\documentclass[modern]{aastex631}
\renewcommand{\ion}[2]{#1\,{\sc #2}}

\shorttitle{Coronal emission measure distribution}
\shortauthors{K. P. Dere}

\graphicspath{{./}{*}}

\begin{document}

\title{The temperature and emission measure distribution in the quiet and active solar corona: a Bayesian approach}

\footnote{Released on January 12, 2022}

\author[0000-0003-1628-6261]{Kenneth P. Dere }
\affiliation{Department of Physics and Astronomy, George Mason University, 4400 University Drive, Fairfax, VA 22030, USA}

\begin{abstract}

The reconstruction of the differential emission measure (DEM) from observations of spectral line intensities provides information on the temperature distribution of the emission measure in the region observed.  The inversion process is known to be highly unstable and it has been necessary to provide additional constraints, such as requiring that the DEM should be smooth.  However, this is a non-physical constraint. The goal of this analysis is to make an empirical determination of the ability of a set of emission line intensities to constrain the reconstruction.  Here, a simple model is used, by means of a Monte-Carlo-Markov-Chain process, to arrive at solutions that reproduces the observed intensities in a region of the quiet Sun and a solar active region.  These solutions are compared by means of the reduced chi-squared.  The conclusion from this analysis is that the observations are only capable of constraining model consisting of 4 temperature-emission measure pairs plus a determination of the standard deviation of the model from the observed line intensities.  A more complex model with 5 temperature-emission measure pairs does not improve the fit and leads to parameters that are irrelevant.  A more general conclusion is that the information content of a set of observed emission lines is limited with respect to the determination of the emission measure distribution.

\end{abstract}

\keywords{Active solar corona --- Quiet solar corona --- Solar EUV emission}

\section{Introduction} \label{sec:intro}

It has been known for some time from ground-based observations that the outer solar atmosphere consisted of a photosphere, chromosphere and an extended corona, visible in eclipse observations.  With the identification of the coronal iron lines \citep{edlen_1943}, the temperature of the corona was appreciated to be quite hot, on the order of 10$^6$ K.

Access to space made it possible to observe the Sun from above the Earth's atmosphere which absorbs the ultraviolet (UV), extreme-ultraviolet (EUV) and X-ray regions of the spectrum.  These latter spectral regions contain continua and spectral lines formed over a broad range of temperatures and provide a detailed picture of the solar atmosphere.  The first ultraviolet spectra were obtained from a V-2 rocket \citep{baum_tousey_1946}.  The early measurements have been summarized by \citet{tousey_1963} and \citet{pottasch}.  In the paper by Pottasch, he goes on to present an analysis of the intensities from a variety of observations spanning the wavelength range from 100 to 1900 \AA.  The primary goal of the Pottasch analysis was to determine the relative abundances of the elements.  In doing this, the emission measure of each ion is determined at the temperature that maximizes the contribution function $G(T)$ (see Equ. \ref{equ:intensity} - \ref{equ:goft}).  These values are multiplied by the relative abundances that provide as little scatter as possible in a plot of the various emission measures as a function of temperature.  His emission measure plot reveals a curve that will become quite familiar, with the minimum of the emission measure occurring near 2 $\times$ 10$^5$ K,  increasing toward chromospheric temperatures and increasing toward coronal temperatures.    \citet{jordan_abund_emd} performed a similar analysis with the XUV spectra reported by \citet{hall_xuv}.  The spectra contained emission lines formed at higher temperatures and the emission measures peaked at a temperature of 1.4 $\times$ 10$^6$ K.

The idea that there exists a continuous differential emission measure (DEM) came to be used in the analysis of solar spectra.  \citet{batstone_1970} used a a DEM consisting of 4 steps to explain their X-ray observations.  \citet{chambe_1971} presented a continuous DEM of an active region that peaked between 1 and 2 $\times$ 10$^6$ K and fell off exponentially at higher temperatures.  \citet{walker_1974} analyzed whole-sun X-ray spectra in terms of a DEM and found a shape similar to that of \citet{chambe_1971}.  \citet{batstone_1970}, \citet{walker_1974} and various other authors have employed either least-squared fits or chi-squared minimization techniques to determine their model parameters.  In the case of \citet{chambe_1971}, the exact process is not clear.

After these initial DEM analyses, it became clear that the derivation of a detailed, smooth DEM had problems.  These were first pointed out by \citet{craig_brown}.  In order to overcome these difficulties, Bayesian techniques have been used to infer the DEM from spectral line observations.  \citet{kashyap_drake} developed a Monte-Carlo-Markov-Chain (MCMC) code with a Metropolis sampler in the Interactive Data Language (IDL).  \citet{warren_bayes} employed a sparse Bayesian analysis to examine the DEM that can be inferred from synthetic spectral line intensities. \citet{hannah_kontar} developed a regularized inversion technique to determine the DEM from observed intensities.  All of these methods apply a smoothing or regularization to the derived DEM.  Here, a different approach is taken to understand the limitations of observed sets of solar emission lines to determine or infer the emission measure structure of the solar atmosphere by means of minimal models requiring few parameters.  No smoothing is applied and the only constraint, other than that provided by the observations, is that the emission measures are positive-definite.

\section{The approach} \label{sec:approach}

The determination of emission measures in the solar atmosphere is based on a number of assumptions about the physical nature of the corona.  They have been stated previously by \citet{kashyap_drake} in their Section 3.1 and essentially repeated here.  The source region is optically-thin for the EUV lines analyzed here, the ionization equilibrium is determined by a balance between collisional ionization and recombination, the ions are excited by collisional excitation by electrons and protons with a Maxwell-Boltzmann distribution, and the elemental abundances do not vary within the source region.  In addition, it is assumed that there is an unknown coronal heating mechanism maintaining the hot corona.

In previous analyses it has generally been assumed that the differential emission measure (DEM) is a continuous, smoothly varying function of temperature.  The approach taken here is to assume that the intensities of the observed spectral line can be reproduced by a set of isothermal emission measures.  \citet{feldman_sumer_isot} and \citet{landi_isot} have demonstrated that for some spectra observed above the limb, the  intensities can be reproduced by a single temperature-emission measure pair (T-EM) , where EM is given by the line of sight integral \mbox{$\int$ N$_e$ N$_H$ d{\it l}}.

Two sets of spectral line observations of \citet{brosius} are examined here.  These observations include a region of the quiet Sun and an active region.  They will be discussed further in Sections \ref{sec:qs} and \ref{sec:ar}.

For each line in the spectrum, the {\it G(T)} function is calculated for a range of temperatures using the atomic parameters in the CHIANTI atomic database \citep{dere_v1, delzanna_v10} with the ChiantiPy software package \citep{ChiantiPy}.  The photospheric elemental abundances of \citet{asplund_abund} as updated by \citet{scott_abund_2, scott_abund_1} are used.  All CHIANTI lines produced by elements having an abundance greater than 10$^{-7}$ that of hydrogen and within 0.05 \AA\ of the observed wavelength are included.  In very few cases is there significant blending.  The temperatures are calculated on a grid with exponential increments of 10$^{0.005}$ K over the temperature range 2.5$\times$10$^5$ to 4.5$\times$10$^6$ for the quiet Sun and 2.5$\times$10$^5$ to 5.6$\times$10$^6$ for the active region.  An average value for the electron density is determined from the available density sensitive lines for each region.  Thus, for a prescribed set of temperatures and emission measures, a spectrum can be calculated.  The question is to find which set of temperatures and emission measures provide the greatest likelihood that the weighted deviations of the observations from the predictions conform to a normal distribution.

A Bayesian approach is followed through the use of a Monte-Carlo-Markov-Chain (MCMC) process to determine the most likely set of temperature-emission (T-EM) pairs that reproduce the observed line intensities.  \citet{kashyap_drake} have provided a good summary of the MCMC technique coupled with the use of the Metropolis sampler \citep{metropolis}.  Here, the MCMC modeling is performed by the open-source Python PyMC3 package \citep{pymc}.  It includes both the Metropolis sampler and the No U-Turn Sampler (NUTS) \citep{NUTS}.  The Metropolis sampler is one of the first that was developed and can be somewhat inefficient since all proposed steps do not lead to a greater likelihood and the proposal process must be repeated, effectively taking a "U-turn" while traversing the Markov chain.  The NUTS sampler was developed for the case of continuous parameter distributions to provide a better and faster sampling algorithm.   The authors list a number of improvements over other samplers, for instance, the ability to efficiently sample with "minimum human interaction."   The model errors are treated as a normal distribution with a mean of zero and a standard distribution as discussed.  The likelihood that is considered by the MCMC chain is the likelihood that the errors are consistent with the assumed normal distribution.  It is necessary to calculate the contribution functions on a grid before doing the MCMC sampling.  With in the framework of the PyMC3 package, it is not possible to recalculate them for each sampling iteration.  At each step, the sampling proposes a set of indices with which the contribution functions can be immediately determined.

The intensity of a spectral line produced by a  single ion is given by 
\begin{equation} 
I\,(i \to f) \, = \,  \frac{h \nu}{4\pi} \, A(i \to f) \, \int N_i \, d{\it l}
\label{equ:intensity}
\end{equation} 
where $h\nu$ is the energy of the emitted photon,  $A(i\to f)$ is the radiative decay rate from initial level {\it i} to final level {\it f}, and $N_i$ is the population density of level {\it i} that is integrated along the line-of-sight {\it l}.  If the energy $h\nu$ is given in erg, then the units of $I(i\to f)$ are in erg~cm$^{-2}$~s$^{-1}$~sr$^{-1}$.

The values of $N_i$ are obtained by solving the equilibrium level balance equations that model the atomic processes populating and de-populating the levels. The atomic parameters include radiative decay rates, electron and proton collisional rate coefficients, autoionization rates, and level-resolved recombination rates from the next-higher ionization state.  The solution of these equations provides the fraction of the ion X$^{+q}$ of the element X in level i.  In the following equation, this is denoted as N$_i$($\rm{X}^{+q}$)/N($\rm{X}^{+q}$).  Once these value are obtained, it is necessary to calculate the value of N$_i$.  This is commonly done by the multiplication of a series of ratios that can be determined.

\begin{equation}
N_i = \frac{N_i(\rm{X}^{+q})}{N(\rm{X}^{+q})} \, \frac{N(\rm{X}^{+q})}{N(\rm{X})} \, \frac{N(\rm{X})}{N_{\rm{H}}} \, N_{\rm{H}}
\end{equation}

Here, N($\rm{X}^{+q}$)/N($\rm{X}$) is the fraction of the ion X$^{+q}$ relative to the element X and is obtained from calculations of the statistical equilibrium between ionization and recombination rates.  N($\rm{X}$)/N$_{\rm{H}}$ is the abundance of the element X relative to hydrogen and is taken from compilations of a variety of measurements.  In very many cases, N$_i$($\rm{X}^{+q}$)/N($\rm{X}^{+q}$) scales linearly with the electron density N$_e$ and it is convenient to divide this term by N$_e$ and then multiply the whole expression by N$_e$.  Then, we arrive at an expression for the predicted line intensity

\begin{equation} 
I\,(i \to f) \, = \, \frac{h \nu}{4\pi} \, A(i \to f) \, \frac{N_i(\rm{X}^{+q})}{N_e \, N(\rm{X}^{+q})} \, \frac{N(\rm{X}^{+q})}{N(\rm{X})} \, \frac{N(\rm{X})}{N_{\rm{H}}} \, \int  N_e N_{\rm{H}} \, d{\it l}
\end{equation}

The G(N$_e$, T) is defined as

\begin{equation}
G(N_e, T) \, = \, \frac{h \nu}{4\pi} \, A(i \to f) \, \frac{N_i(\rm{X}^{+q})}{N_e \, N(\rm{X}^{+q})} \, \frac{N(\rm{X}^{+q})}{N(\rm{X})} \, \frac{N(\rm{X})}{N_{\rm{H}}}
\end{equation}

and 

\begin{equation}
I\,(i \to f) \, = \, G(N_e, T) \, \int N_e N_{\rm{H}} \, d{\it l}
\label{equ:goft}
\end{equation}

where the line-of-sight emission measure is $\int N_e N_{\rm{H}} \, d{\it l}$.

For the present analyses, the emission measure distribution is given by

\begin{equation}
EM(T) = \sum_i \delta(T - T_i) EM_i 
\end{equation}

For each T-EM pair, there are two variables, the index for the temperature array and the EM.  The temperature indices i$_n$ are constrained such that
\begin{equation}
i_{min} \leq i_0 < i_1 < i_2 \cdots i_m \leq i_{max} 
\end{equation}
where i$_{min}$ is the minimum index for the temperature array, usually zero, and i$_n$ are the temperature indices where n = 0 through the number of temperatures less one.  The maximum index i$_{max}$  is the largest index value of the temperature array.  The temperature indices are sampled with a Metropolis sampler available in PyMC3 and are modeled as a discrete-uniform distribution provided by PyMC3.  The EM values are constrained to be positive-definite by the use of the logarithm of the EM as the sampling variable.  Each value of EM$_i$ is modeled as a continuous-uniform distribution, also provided by PyMC3, where minimum and maximum values are set.  The values of EM$_i$ are sampled with the NUTS sampler \citep{NUTS}.  The uniform prior distributions are relatively uninformed but maximum and minimum values can be set for the temperatures and each EM$_i$ and a starting point for the sampling provided.

The model consists of a prescribed number T-EM pairs.  Sequential models are constructed, beginning with 2 pairs.  Initial values of the T-EM are estimated from the {\it em-loci} plots such as displayed in Figs. \ref{fig:qs_emplot} and \ref{fig:ar_emplot}.  After sufficient tuning steps, the MCMC samplers are run to provide a posterior distribution of the temperature and EM values that best reproduce the observed spectral intensities.  From the posterior distributions, the means are determined for each value of T and EM and inserted back into the model to predict the spectrum.  A value of $\chi^2$ can then be calculated for each set of T-EM pairs.  
\begin{equation}
\chi^2 = \sum_i ((I_i - P_i)/\sigma_i)^2
\label{equ:chi-squared}
\end{equation}
where I$_i$ is the observed intensity for line i, P$_i$ is the predicted intensity and $\sigma_i$ is an estimate of the combined error in observed and predicted intensities.  As in \citet{dere_serts_densities}, an estimate of the standard deviation $\sigma = w \times I_i$ is used.  In \citet{dere_serts_densities} a value of $w$ = 0.2 was determined for comparing lines of the same ion.  Since the appropriate value of $w$ is not known beforehand, it has been necessary to determine it in an iterative manner.  First, a 2 T-EM model with an estimate for $w$ was used and evaluated by means of MCMC modeling.  The values of the T-EM pairs are determined from the mean of the posterior distributions of each of the parameters.  These values are then used to predict the spectral line intensities and the standard deviation of the difference between the predictions and the observations. This value of the standard deviation determines a new value of $w$.  The 2 T-EM model is run again with this new value of $w$ to determine an improved estimate of $w$.  This process is then repeated with the 3, 4, and 5 T-EM models in that order.  A final estimate of $w$ = 0.3 is arrived at with the 4 and 5 T-EM models.  All models are then run with this final value of $w$ = 0.3.  In \citet{dere_serts_densities} a value of $w$ = 0.2 was determined.  The difference can be explained by the fact that in \citet{dere_serts_densities} only intensities of lines of the same ion were compared.  Consequently, whatever errors that are caused by uncertainties in the elemental abundances or ionization equilibrium are absent.  \citet{brosius} lists uncertainties for their intensity measurements.  These are roughly half of the value of 0.3 used here that accounts for the combined uncertainties in the intensity measurements, the atomic data and the model.

The procedure is to iterate over the number of T-EM pairs, starting with 2 pairs and continuing to greater numbers of pairs, using the final value of $w$ = 0.3.  The results for each pair is compared by means of the reduced chi-squared \citep{bevington}.
\begin{equation}
\chi^2_{\nu} = \chi^2 \, / \, \nu
\label{equ:reduced_chisq}
\end{equation}
where $\nu$ is the degrees of freedom N$_{obs}$ - N$_{params}$, N$_{obs}$ is the number of observations and N$_{params}$ is the number of parameters in the model.  The number of parameters is 2 $\times$ the number of T-EM pairs plus 1.  The fact that it has been necessary to determine the value of $w$ accounts for the additional parameter.  The value of N$_{obs}$ is taken as the number of observed spectral lines.  
Values of $\chi^2_{\nu}$ should be about 1 if the data and the model are appropriate.  "Values of $\chi^2_{\nu}$ much larger than 1 result from large deviations from the assumed distribution and may indicate poor measurements, incorrect assignment of uncertainties, or an incorrect choice of probability function.  Very small values of $\chi^2_{\nu}$ are equally unacceptable and may imply some misunderstanding of the experiment \citep{bevington}".

\section{The analysis of a quiet sun spectrum} \label{sec:qs}

The observed spectral line intensities are taken from Table 2 of \citet{brosius} and were obtained in a quiet region of the Sun in 1993. Spectral line observations were made in a wavelength range between 274 and 417 \AA\ and include 57 spectral lines formed by 19 ions.   The intensities tabulated by \citet{brosius} were obtained by averaging over the 282 arc-sec slit. The quoted spatial resolution is about 5 arc-sec.     

Density sensitive line ratios in the data set have been analyzed by \citet{dere_serts_densities}.  Line intensity ratios from ions of \ion{Mg}{viii}, \ion{Si}{ix}, \ion{Si}{x}, \ion{Fe}{xi}, \ion{Fe}{xii},  \ion{Fe}{xiii}, and \ion{Fe}{xiv} were used to derive an average electron density of 5 $\times$ 10$^{8}$ cm$^{-3}$.  Of these ions, the most robust results were obtained with \ion{Si}{x} and 
\ion{Fe}{xii}, \ion{Fe}{xiii}, and \ion{Fe}{xiv}.  The analysis used both a chi-squared minimization scheme and MCMC sampling of the electron density-EM space at a temperature specific to each ion.  There two methods provided a reasonable agreement with each other.

Most of the quiet sun lines listed by \citet{brosius} were included in the present analysis, with the exception of \ion{He}{ii} $\lambda$ 304.  It is formed at a much lower temperature than the other lines and would likely require its own T-EM pair that would be indeterminate.  In addition, one of the line identifications were updated.  The line at 359.374 \AA\ identified by \citet{brosius} as \ion{Ne}{v} (2s$^2$\,2p$^2$ $^3$P$_2$  - 2s\,2p$^3$ $^3$S$_1$) is found to be blended with a much stronger \ion{Ca}{viii} line (3s$^2$\,3p $^2$P$_{3/2}$ - 3s$^2$\,3d $^2$D$_{5/2}$).  This is true both in the quiet Sun spectra and the active region spectra discussed later in Section \ref{sec:ar}.  The final list consists of 53 spectral lines produced by 19 ions.  The temperature grid consisted of 271 exponentially spaced increments of 10$^{0.005}$ K from 2.5 $\times$ 10$^5$ to 4.5 $\times$ 10$^6$ K.  The {\it G(T)} functions for each line, including possible blends, were calculated over this temperature range using the atomic data in Version 10 of the CHIANTI database \citep{delzanna_v10}.

\begin{figure}[ht!]
\plotone{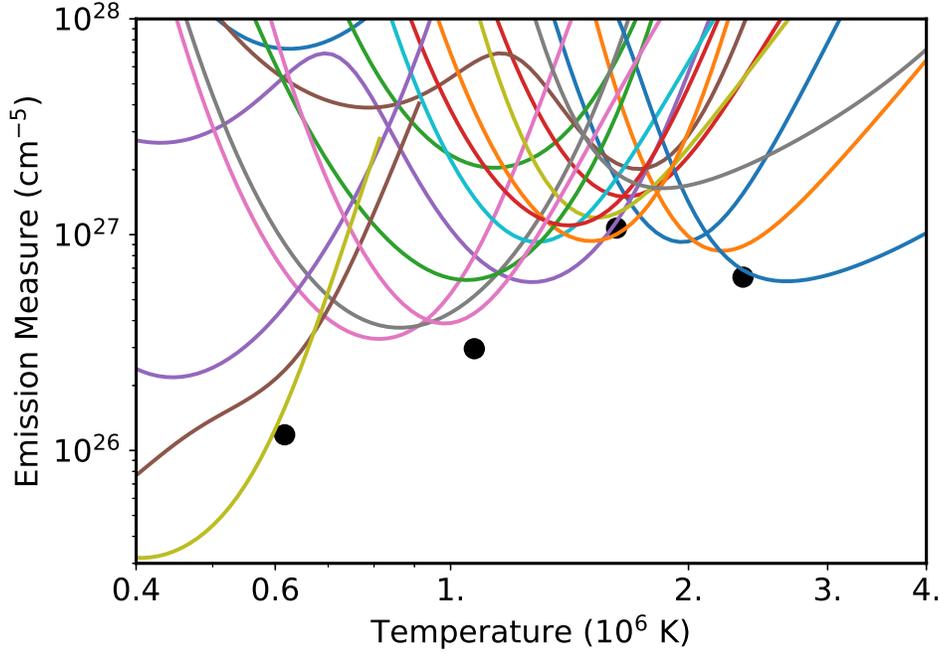}
\caption{The EM-{\it loci} for the abbreviated list of spectral lines in the 1993 quiet Sun spectra.  In addition, the values for the four T-EM pair final solution (dots).  \label{fig:qs_emplot}}
\end{figure}

For the purposes of displaying the EM-{\it loci} of this data set, an abbreviated spectral line set was created with only a single line per ion.  The selection process was very straightforward with the shortest wavelength line of each ion selected, basically, the first in the list.  These are shown in Fig. \ref{fig:qs_emplot}.  Also, the final solution, a four T-EM pair model, is displayed as the solid dots in the figure.

Once the appropriate standard deviation parameter $w$ has been determined, as discussed above in $\S$ \ref{sec:approach}, the determination of the most likely solutions proceeds.  By starting with a 2 T-EM pair model, it is first necessary to find prior estimates of the 2 temperatures and 2 emission-measures.  This is done by visually inspecting the EM-{\it loci} plots displaying all of the EM-{\it loci} curves, considerably more than in Fig. \ref{fig:qs_emplot}, to find where there are many EM-{\it {\it loci}} curves that form a dense pattern, following \citet{feldman_sumer_isot, landi_isot}.  In some cases the estimate of the prior parameters results in a posterior distribution that is located at a distance from the prior values.  However, a very poor choice of the prior starting point can lead to an unrealistic outcome.  Following the analysis with the 2 T-EM model a similar analysis is performed with the 3 T-EM model with improved information about the priors obtained from the 2 T-EM modeling.  This is followed by application of the 4 and 5 T-EM models.  For the larger number of T-EM pairs, the prior estimates for the temperature are often set at equally spaced values between reasonable minimum and maximum values.  These latter models provide the best estimates of the standard deviation of the observed intensities with respect to the predicted intensities.

\begin{figure}[ht!]
\plotone{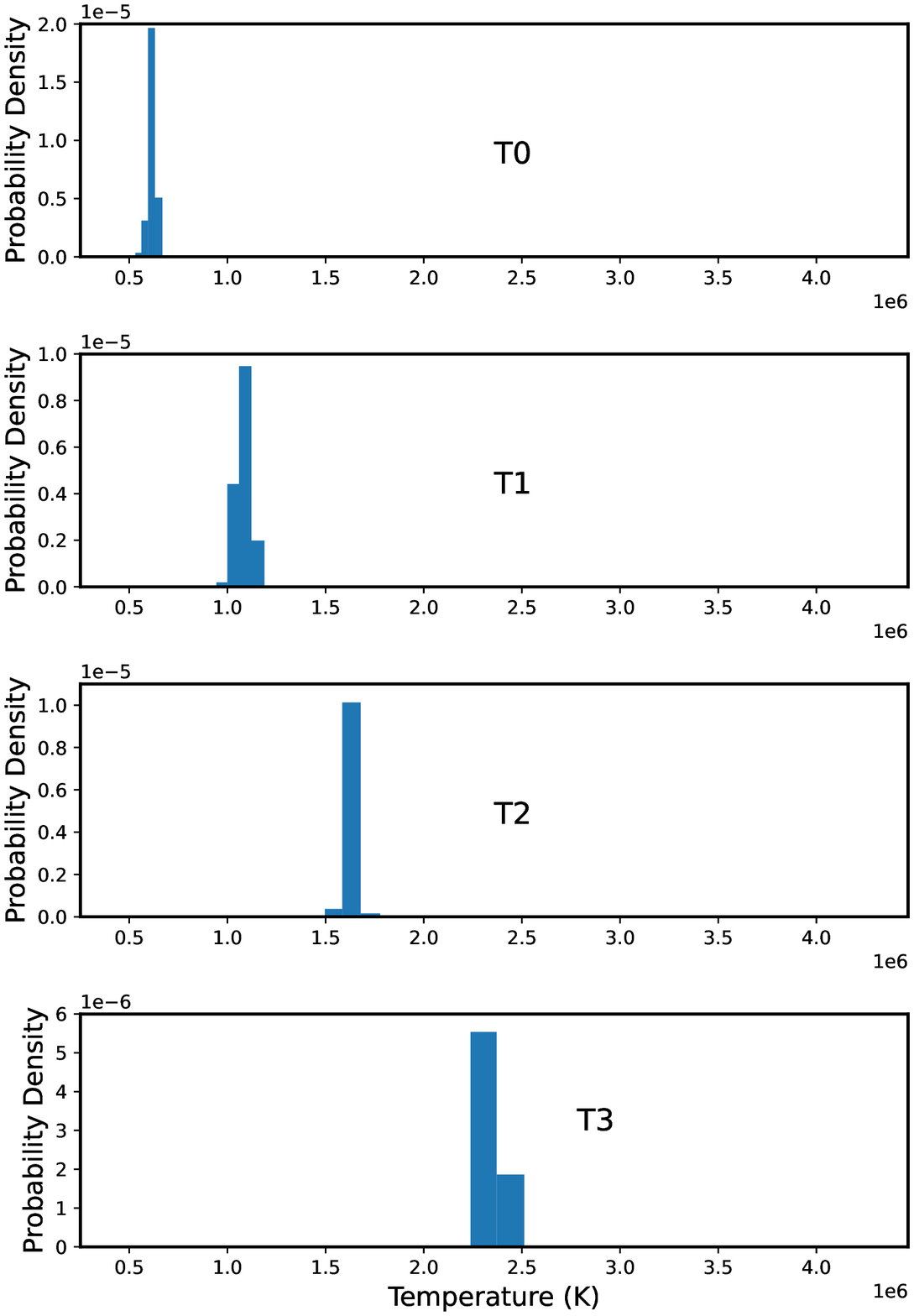}
\caption{The posterior probability density for the temperature of the 4 T-EM quiet Sun model.  \label{fig:qs_t_stackplot}}
\end{figure}

\begin{figure}[ht!]
\plotone{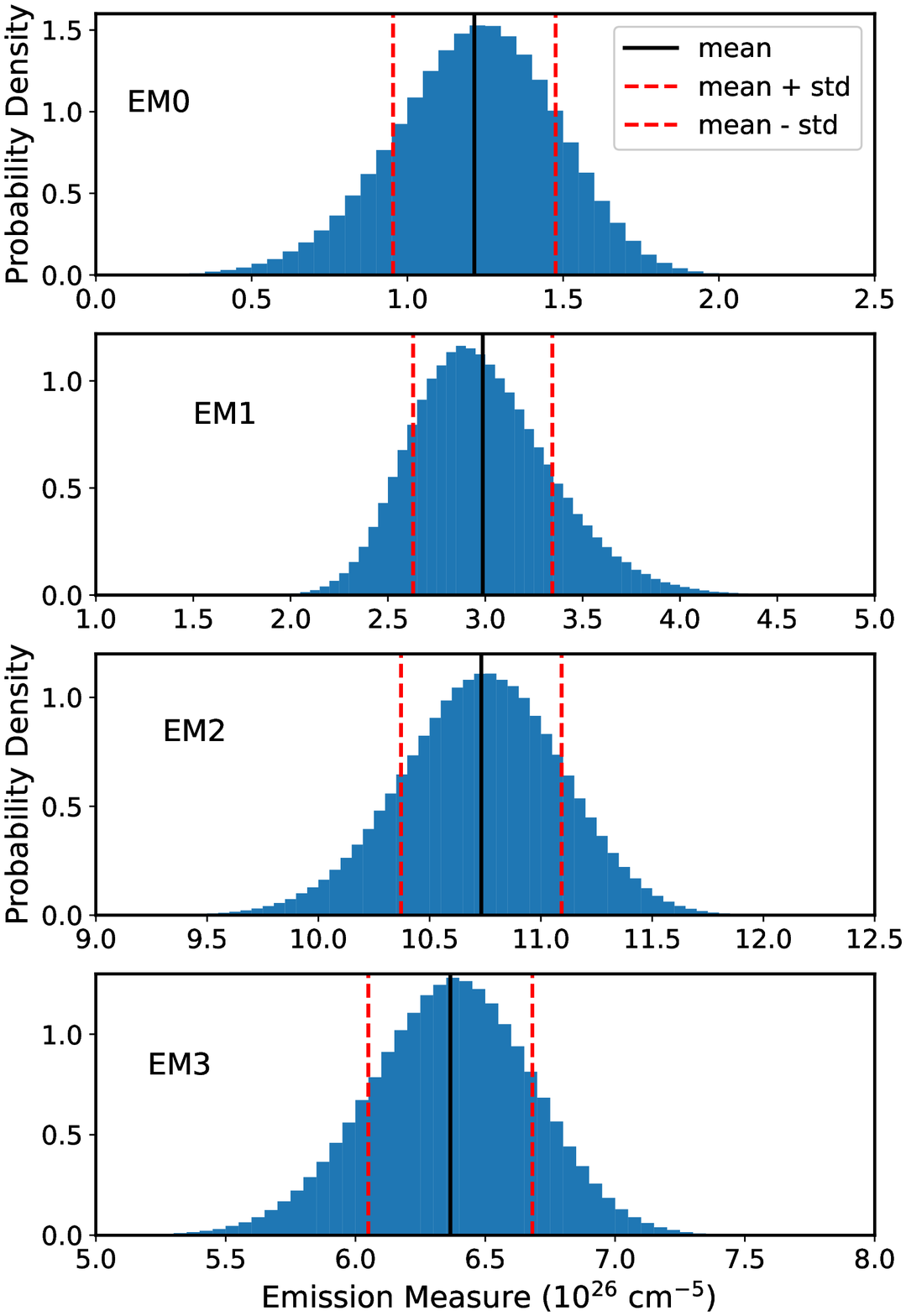}
\caption{The posterior probability density for the EM of the 4 T-EM quiet Sun model.  \label{fig:qs_em_stackplot}}
\end{figure}

The process of evaluating 2, 3, 4, and 5 T-EM pair models was then repeating but with a common value of $w$ = 0.3.  For each evaluation of the models, the best parameters are determined for each T-EM model pair.  The posterior distributions for the 4 T-EM model for the temperatures  are shown in Fig. \ref{fig:qs_t_stackplot} and the posterior distributions for the EM are shown in Fig. \ref{fig:qs_em_stackplot}.  From these posterior distributions, the values for the 4 T-EM pairs are derived as the mean and standard deviations.  These values are presented in Table \ref{tab:qs_T-EM}.  The values of the standard deviation of the derived posterior parameters are quite small.  If they were to be plotted in Fig. \ref{fig:qs_emplot}, they would be almost invisible.

\begin{figure}[ht!]
\plotone{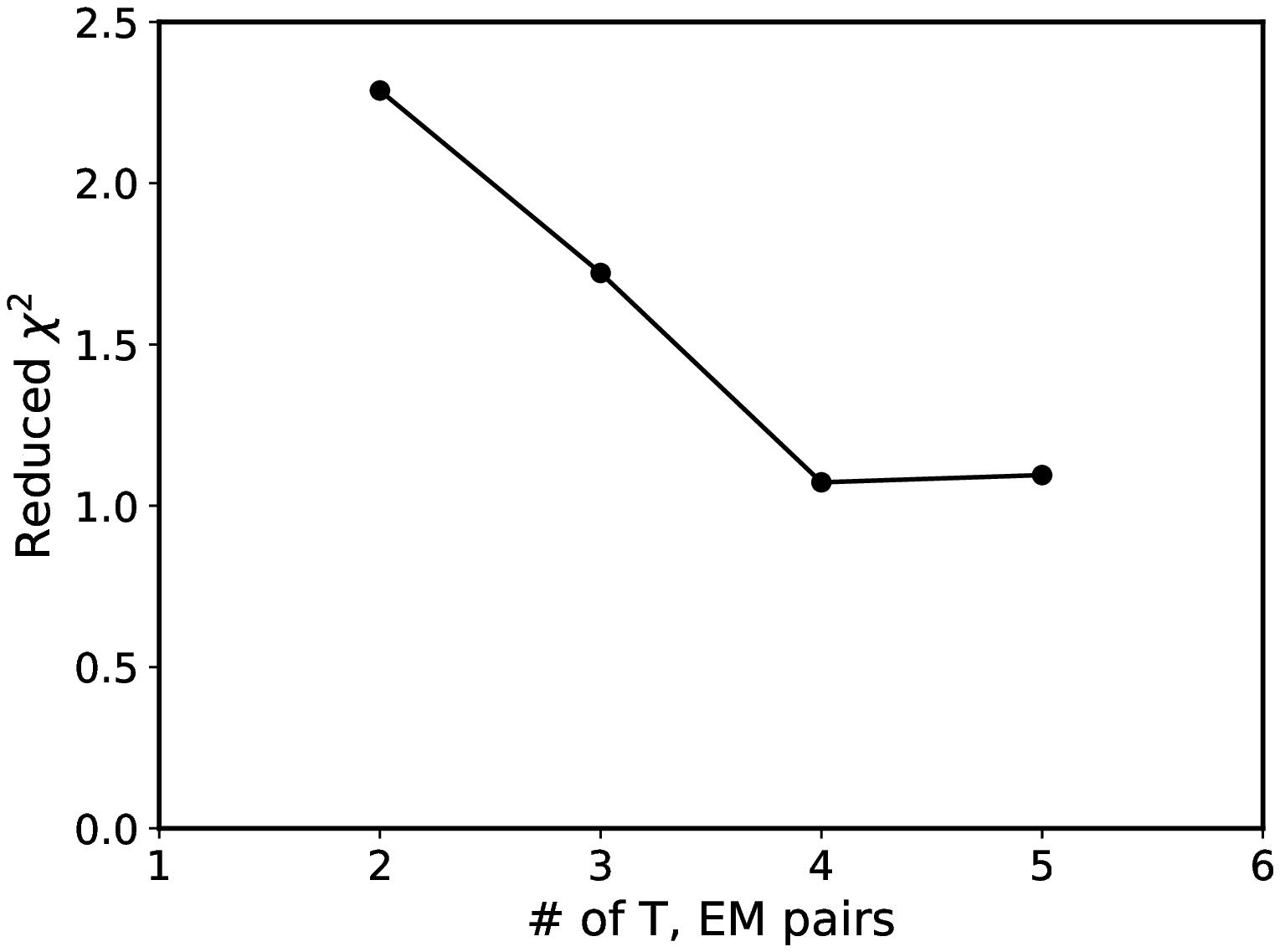}
\caption{The values of $\chi^2_{\nu}$ as a function of the number of T-EM pairs in each quiet Sun model.  \label{fig:qs_reduced_chisq}}
\end{figure}

These derived parameters, consisting of the mean value of the posterior distributions, are then used to calculate a predicted spectrum and the value of $\chi^2$, following Equ. \ref{equ:chi-squared}.  The values found for $\chi^2$ are 110 for the 2 T-EM pairs model, 79 for the 3 T-EM pairs model, 47 for the 4 T-EM pairs model, and 46 for the 5 T-EM pairs model.  The values of $\chi^2_{\nu}$ (Equ. \ref{equ:reduced_chisq}) are shown in Fig. \ref{fig:qs_reduced_chisq} {\it versus} the number of T-EM pairs.  From this figure it can be seen that using 5 T-EM pairs does not improve the fit over the use of 4 T-EM pairs.  In fact, the solution for the 5 T-EM pairs has one T-EM pair where, after sufficient tuning with the MCMC procedure, the value of the EM is so low that it does not effectively contribute to the predicted spectrum.

\begin{table}[ht!]
\begin{center}
\caption{ Temperature, Emission Measures and their standard deviations ($\sigma$) for the 4 T-EM quiet Sun model}
\label{tab:qs_T-EM}
\begin{tabular}{cc}
\hline
T $\pm$ $\sigma$ (10$^6$ K) & EM $\pm$ $\sigma$ (10$^{26}$ cm$^{-5}$) \\
\hline
0.612  $\pm$ 0.0018  &   1.21 $\pm$ 0.26 \\
1.08   $\pm$ 0.036  &   2.99 $\pm$ 0.36 \\
1.62   $\pm$ 0.024 &   10.7  $\pm$ 0.36 \\
2.33   $\pm$ 0.045 &     6.37 $\pm$ 0.32 \\
\hline
\end{tabular}
\end{center}
\end{table}

\begin{figure}[ht!]
\plotone{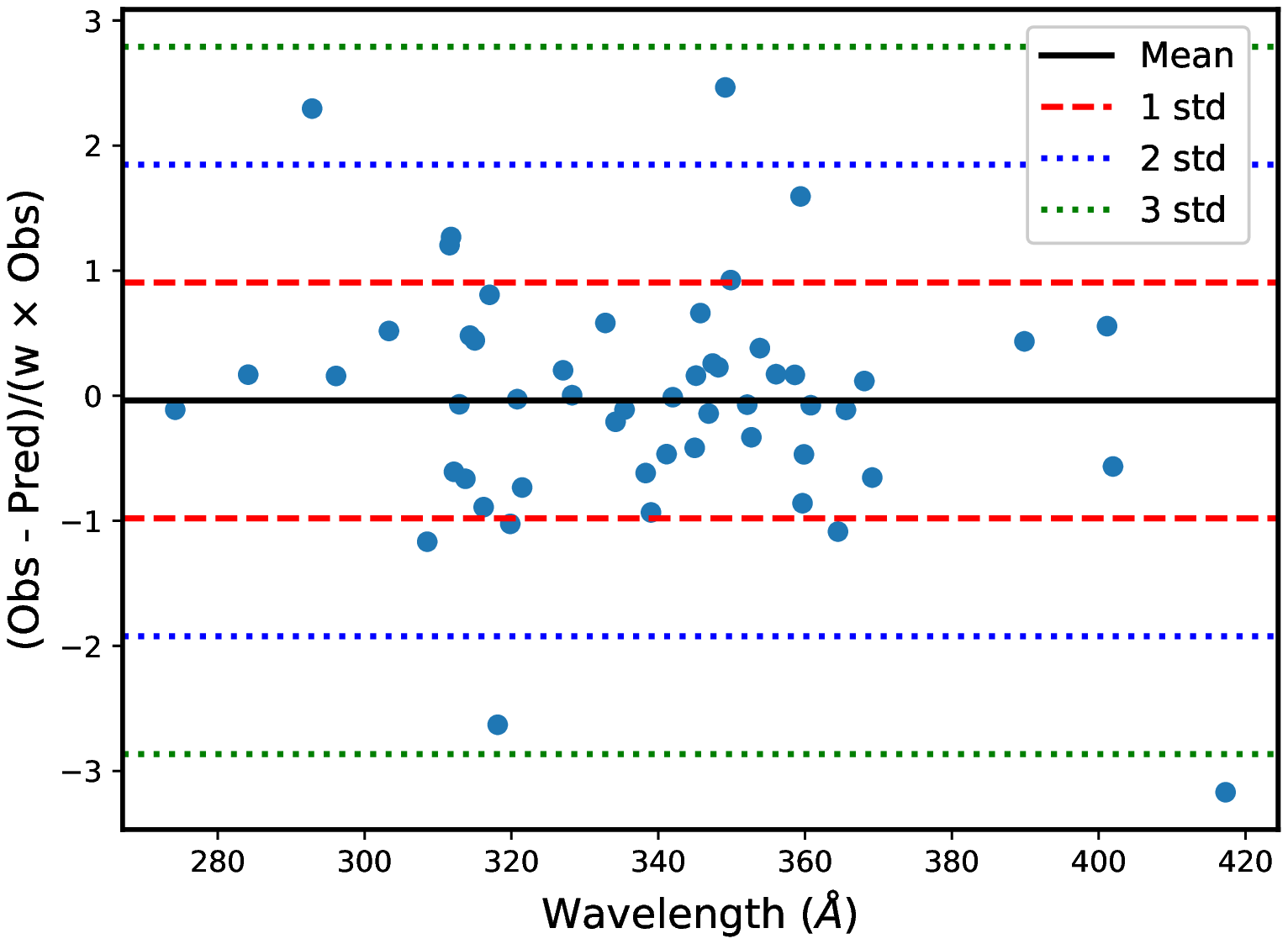}
\caption{The values of the weighted deviations for the quiet Sun spectra {\it vs} wavelength.} \label{fig:qs_weighted_dev}
\end{figure}

In Fig. \ref{fig:qs_weighted_dev} the weighted deviations $(I_i - P_i)/\sigma_i$ together with the average and 1, 2 and 3 standard deviations (std) are shown.  The values are roughly consistent with a normal distribution with 77\% being within 1 std, 92\% within 2 std and 98\% within 3 std, compared with the values for the normal distribution of 68\%, 95\% and 99.7\%, respectively.

\section{The analysis of an active region spectrum} \label{sec:ar}

The observed spectral line intensities are taken from Table 1 of \citet{brosius} and were obtained in a solar active region in 1993. Spectral line observations were made in a wavelength range between 274 and 417 \AA\ and include 65 spectral lines formed by 25 ions. The spectral lines are formed over a temperature range from 5 $\times$ 10$^5$ to 1.6 $\times$ 10$^7$ K, \citep{brosius}. The intensities tabulated by \citet{brosius} were obtained by averaging over the 282 arc-sec slit. The quoted spatial resolution is about 5 arc-sec.

\subsection{Electron densities in the active region} \label{subsec:ne}

\begin{figure}[ht!]
\plotone{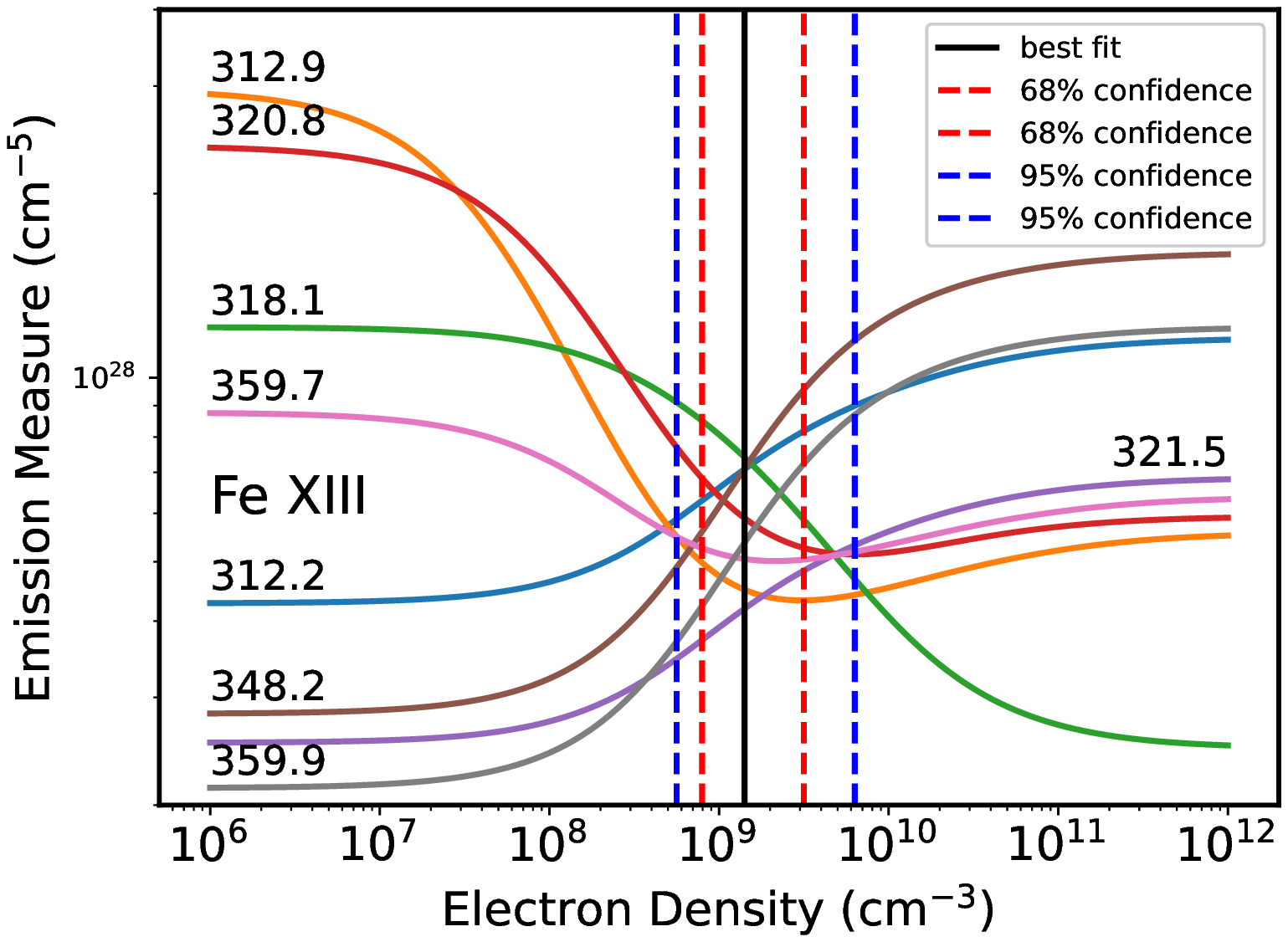}
\caption{The EM-{\it loci} for the \ion{Fe}{xiii} lines for the active region used in the analysis {\it vs} electron density.  The minimum of $\chi^2$ occurs at an electron density of 1.4$\times$ 10$^9$ cm$^{-3}$.  The regions for a statistical confidence of 68\% and 95\% are also shown. \label{fig:ar_fe_13_emplot_best_68_95_r3.eps}}
\end{figure}

\begin{figure}[ht!]
\plotone{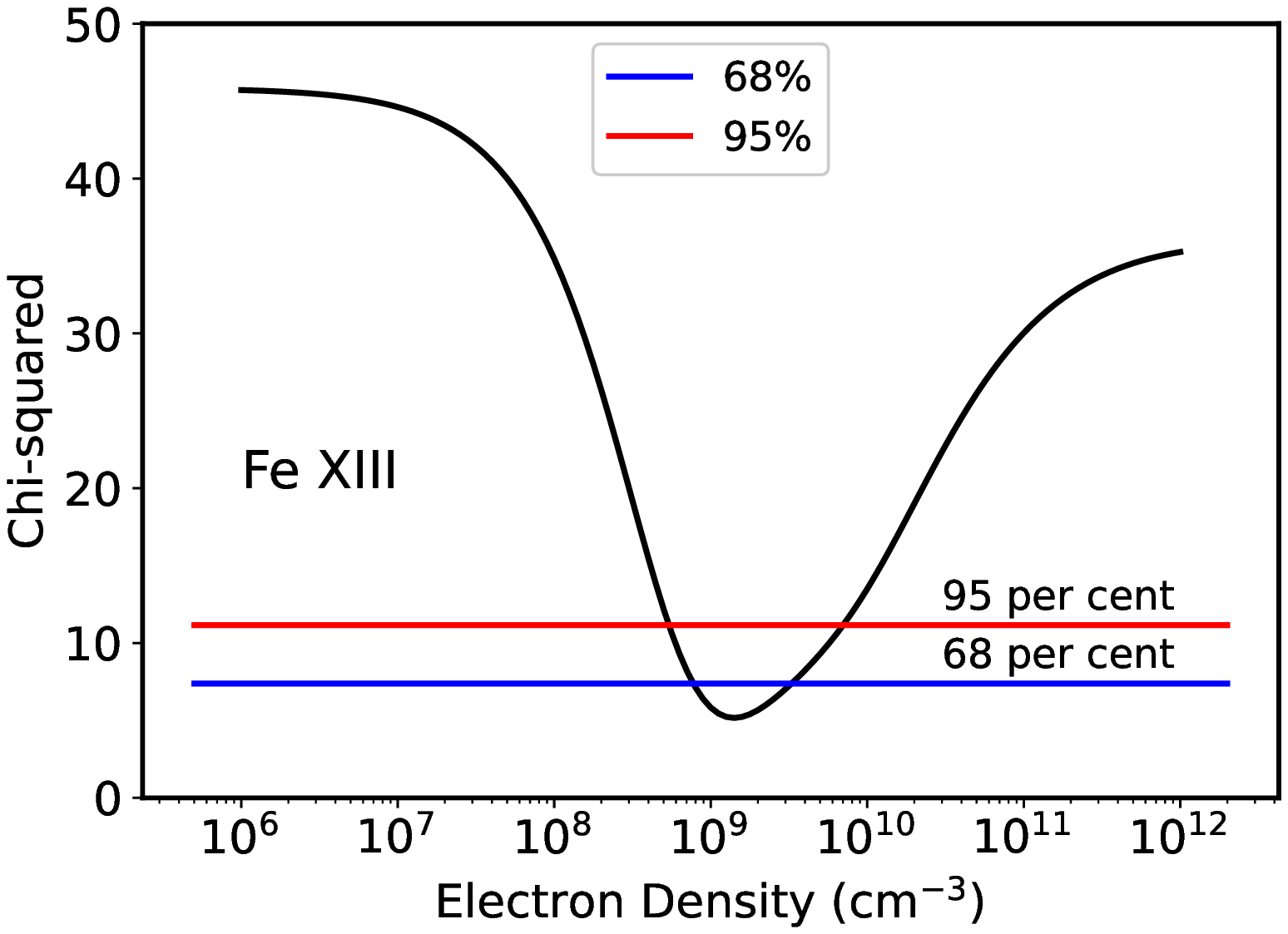}
\caption{$\chi^2$ as a function of electron density as a function of electron density for \ion{Fe}{xiii} for the active region.  The minimum of $\chi^2$ occurs at an electron density of 1.4$\times$ 10$^9$ cm$^{-3}$.  \label{fig:ar_fe_13_chisq_vs_ne.eps}}
\end{figure}

The electron densities in the active region have been derived by means of density-sensitive line-ratios of the ions \ion{Fe}{xi}, \ion{Fe}{xii},  \ion{Fe}{xiii}, and \ion{Fe}{xiv}.  The determination of the densities has been performed by a straightforward $\chi^2$ minimization process as used by \citet{dere_serts_densities}.  The observed intensities of each line of the diagnostic ion are divided by their contribution function {\it G(T)} as a function of electron density at the temperature where the {\it G(T)} function peaks.  For \ion{Fe}{xiii} this temperature is 1.78 $\times$ 10$^6$ K.  The EM-{\it loci} for the \ion{Fe}{xiii} lines are shown in Fig. \ref{fig:ar_fe_13_emplot_best_68_95_r3.eps}.  In addition, the density that provides the best fit, as well as, the regions of 68\% and 95\% statistical confidence are also shown in Fig. \ref{fig:ar_fe_13_chisq_vs_ne.eps}.  The regions of confidence are found by using the prescription of \citet{lampton}.  This method was found to be comparable to a determination of these quantities through an MCMC analysis of the quiet Sun density diagnostics \citep{dere_serts_densities}.

\begin{figure}[ht!]
\plotone{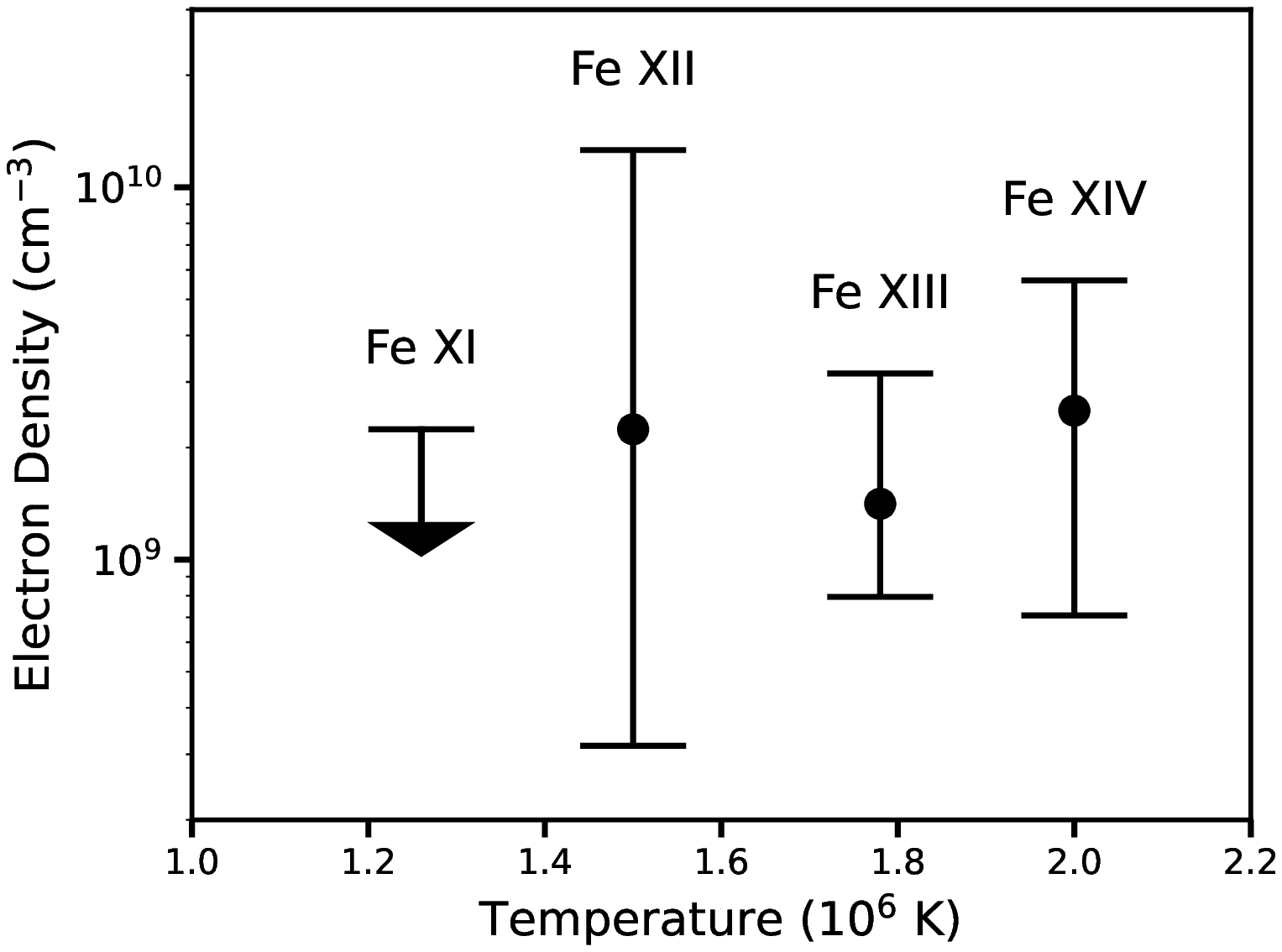}
\caption{Electron densities derived from the lines of \ion{Fe}{xi}, \ion{Fe}{xii}, \ion{Fe}{xiii}, and \ion{Fe}{xiv} as a function of the temperature of their peak emissivity for the active region.  These are compatible with an average electron density of 2$\times$10$^9$ cm$^{-3}$.  \label{fig:ar_densities.eps}}
\end{figure}

The same procedures used to derive electron densities from the \ion{Fe}{xiii} lines have also been applied to the lines of \ion{Fe}{xi}, \ion{Fe}{xii}, and \ion{Fe}{xiv}.  The results of this analysis are displayed in Fig. \ref{fig:ar_densities.eps} where the best fit densities as well as the regions of 68\% confidence are also shown.  For \ion{Fe}{xi} only an upper limit can be obtained.  From these results, an average electron density of  2$\times$10$^9$ cm$^{-3}$ is chosen.

\citet{brosius} found somewhat higher densities of 5 $\times$ 10$^9$ cm$^{-3}$ in their 1993 active spectra.  As suggested by \citet{dere_serts_densities}, the difference is largely due to the fact the only distorted wave calculations were available to provide the calibration of densities {\it vs} line ratios.  These tend to underestimate some of the important excitation rates responsible for the dependence of the level populations on the electron density.

\subsection{The emission measure distribution in the active region} \label{subsec:ar-emd}

\begin{figure}[ht!]
\plotone{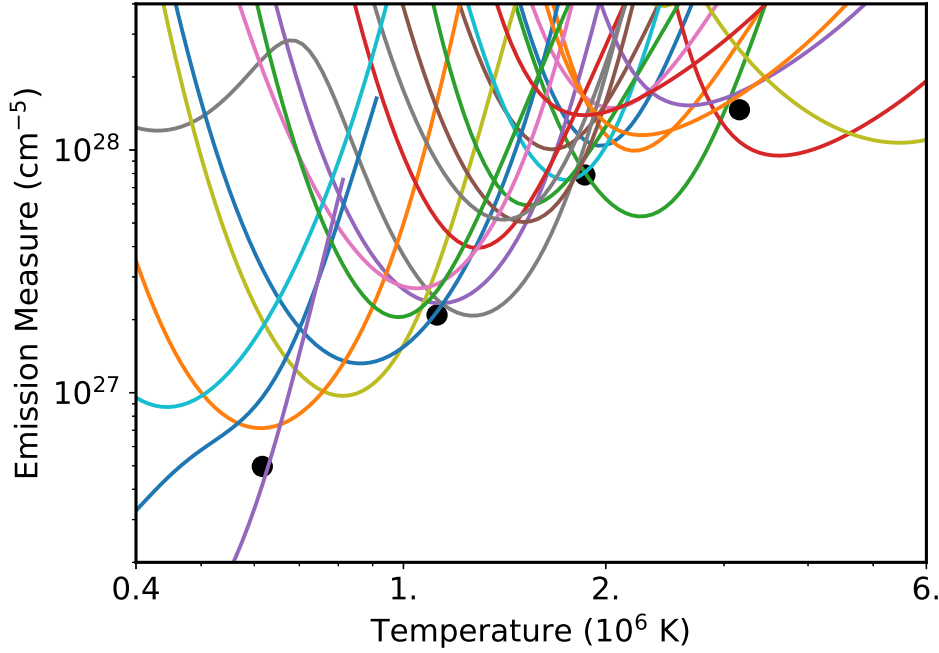}
\caption{The EM-{\it loci} for the abbreviated list of lines in the 1993 active region spectra.  In addition, the values for the four T-EM pair final solution (dots).  \label{fig:ar_emplot}}
\end{figure}

The temperature grid consists of 271 temperatures exponentially spaced by increments of 10$^{0.005}$ K from 2.5 $\times$ 10$^5$ to 5.6 $\times$ 10$^6$ K.  Again, not all of the 65 spectral lines in the \citet{brosius} spectrum were used.  The final list consists of 63 spectral lines formed by 25 ions.  The emissivities of each line are calculated at an electron density of 2$\times$10$^9$ cm$^{-3}$, otherwise, the procedures are the same as followed in Sec. \ref{sec:qs} for the quiet Sun.  An abbreviated set of EM-{\it loci} for the active region is displayed in Fig. \ref{fig:ar_emplot}.  Also, the final solution for a four T-EM pair model, is displayed as the solid dots in the figure.

\begin{figure}[ht!]
\plotone{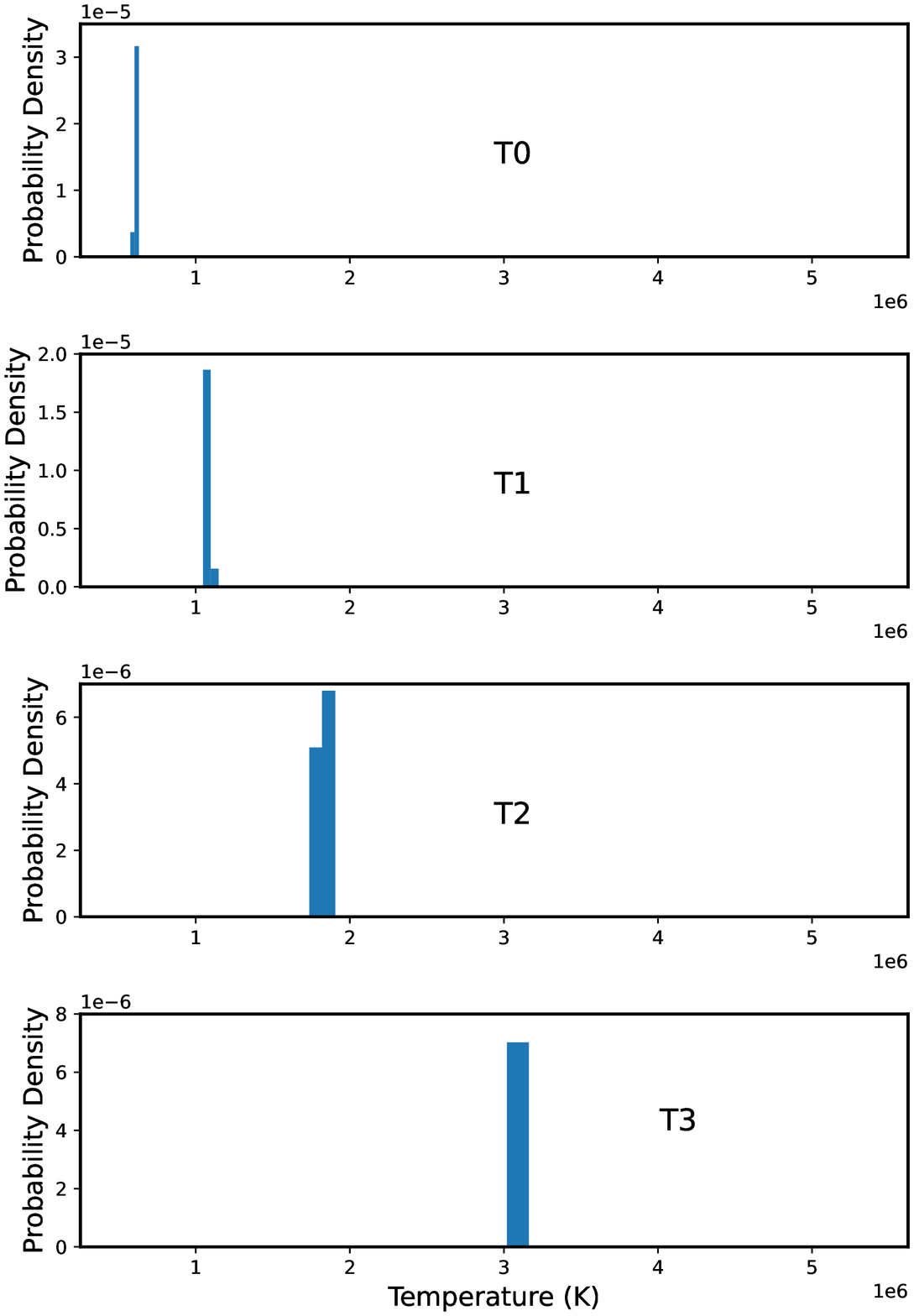}
\caption{The posterior probability density for the temperature of the 4 T-EM active region model.  \label{fig:ar_t_stackplot}}
\end{figure}

\begin{figure}[ht!]
\plotone{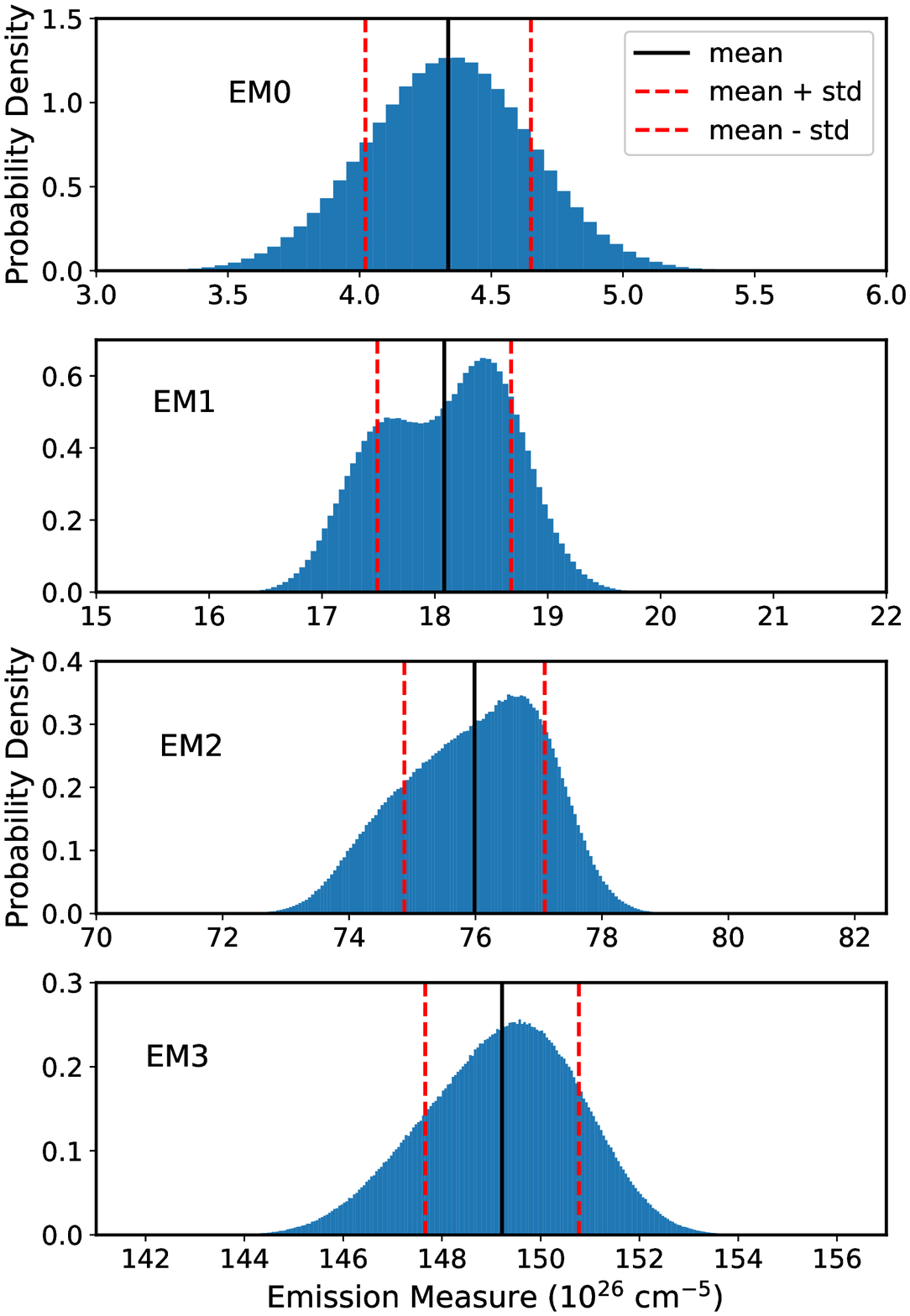}
\caption{The posterior probability density for the EM of the 4 T-EM active region model.  \label{fig:ar_em_stackplot}}
\end{figure}

\begin{figure}[ht!]
\plotone{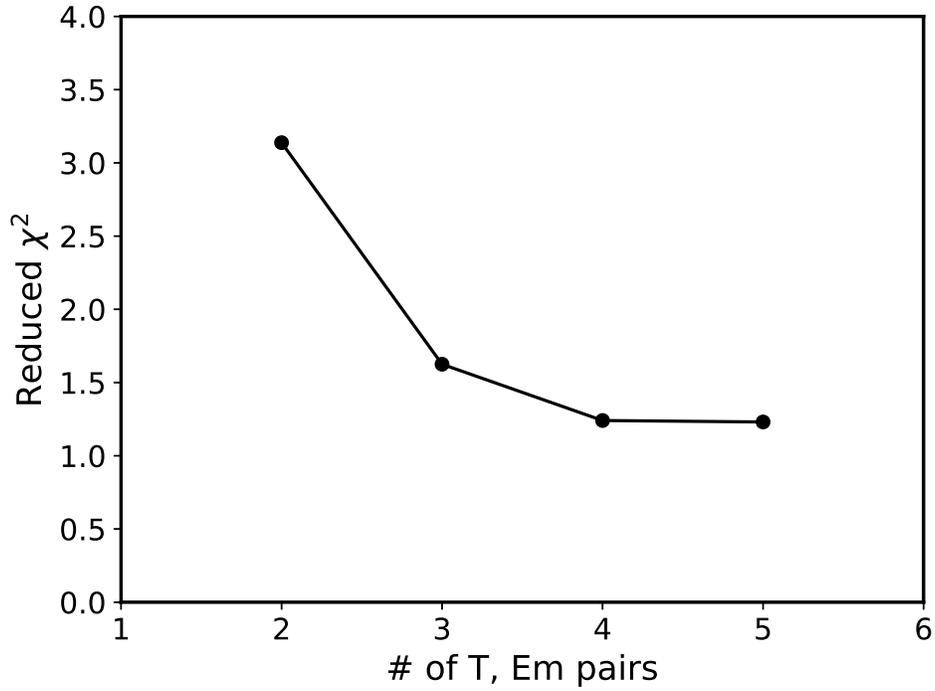}
\caption{ The values of $\chi^2_{\nu}$ as a function of the number of T-EM pairs in each model of the active region. \label{fig:ar_reduced_chisq}}
\end{figure}

For the case of the best 4 T-EM pair model, the posterior probability densities for the temperatures are shown in Fig. \ref{fig:ar_t_stackplot} and those for the emission-measures are shown in Fig. \ref{fig:ar_em_stackplot}. The derived parameters, consisting of the mean value of the posterior distributions for temperature and emission measure, are then used to calculate a predicted spectrum and the value of $\chi^2$ is calculated, following Equ. \ref{equ:chi-squared} for each of the T-EM pair models.  The values found for $\chi^2$ is 182 for the 2 T-EM pairs model, 91 for the 3 T-EM pairs model, 66 for the 4 T-EM pairs model, and 64 for the 5 T-EM pairs model.  The values of $\chi^2_{\nu}$ (Equ. \ref{equ:reduced_chisq}) are shown in Fig. \ref{fig:ar_reduced_chisq} with the number of observations N$_{obs}$ equal to the number of lines observed.  From this figure it can be seen that using 5 T-EM pairs does not improve the fit over the use of 4 T-EM pairs.  The solution for the 5 T-EM pairs has one T-EM pair that does not contribute much to the predicted spectrum.  The parameter values for the 4 T-EM model are provided in Table \ref{tab:ar_T-EM}.  The values of the standard deviation are again quite small.

\begin{table}[ht!]
\begin{center}
\caption{Temperature, Emission Measures and their standard deviations ($\sigma$) for the 4 T-EM active region model}
\label{tab:ar_T-EM} 
\begin{tabular}{cc}
\hline
T $\pm$ $\sigma$ (10$^6$ K) & EM $\pm$ $\sigma$ (10$^{26}$ cm$^{-5}$) \\
\hline
0.608  $\pm$ 0.0068  &   4.34 $\pm$ 0.31 \\
1.08   $\pm$ 0.0011  &   18.1 $\pm$ 0.59 \\
1.81   $\pm$ 0.01 &   76.0  $\pm$ 0.11 \\
3.07   $\pm$ 0.03 &    149. $\pm$ 1.5 \\
\hline
\end{tabular}
\end{center}
\end{table}

\begin{figure}[ht!]
\plotone{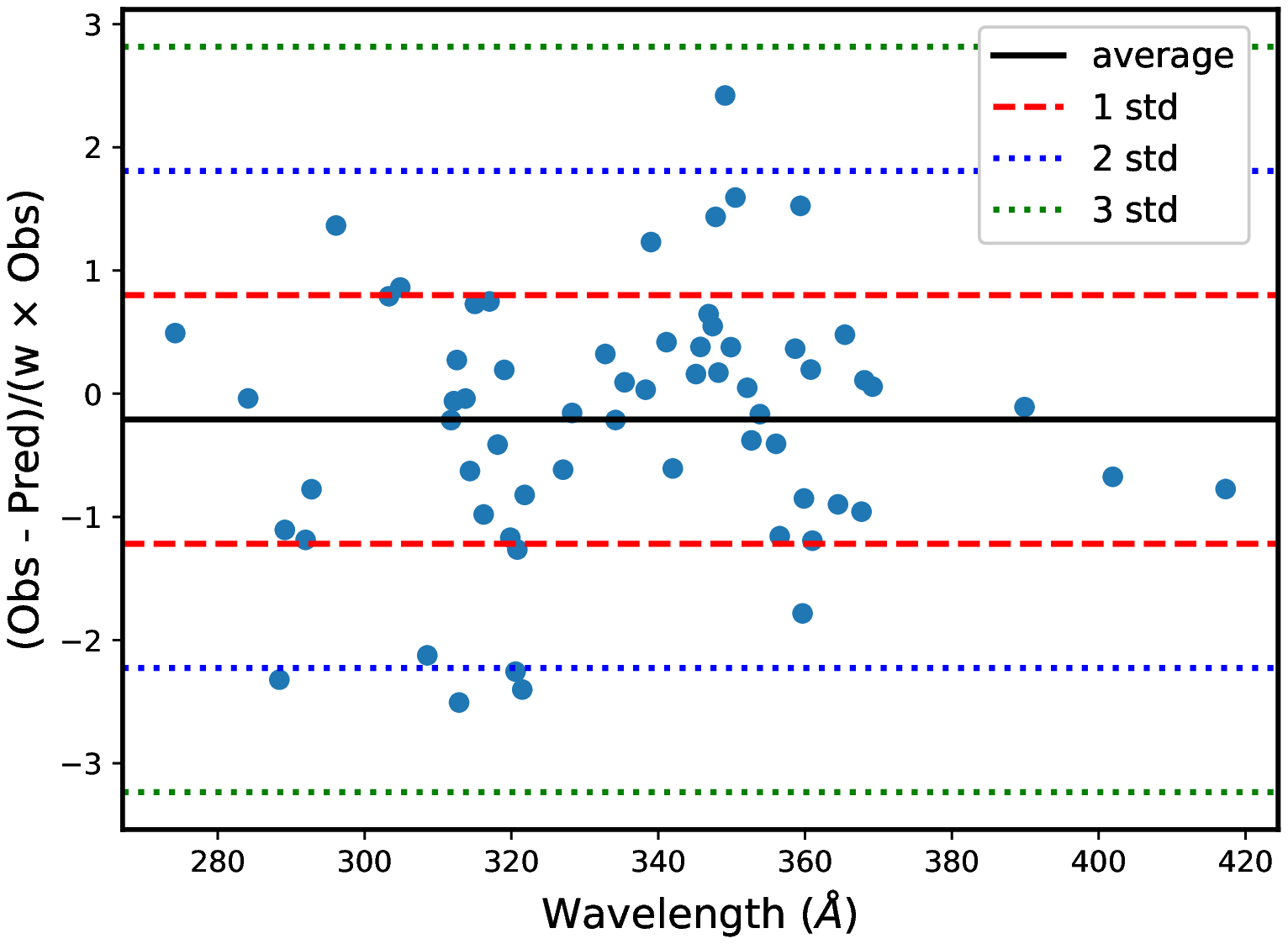}
\caption{The values of the weighted deviations for the active region spectra {\it vs} wavelength.} \label{fig:ar_weighted_dev}
\end{figure}

In Fig. \ref{fig:ar_weighted_dev} the weighted deviations $(I_i - P_i)/\sigma_i$ together with the average and 1, 2 and 3 standard deviations (std) shown for the 1993 active region spectra.  The values are roughly consistent with a normal distribution with 62\% being within 1 std, 90\% within 2 std and 100\% within 3 std, compared with the values for the normal distribution of 68\%, 95\% and 99.7\%, respectively.

\section{Discussion} \label{sec:discussion}

The DEM for these spectra have been previously determined by \citet{brosius} and \citet{kashyap_drake}.  In the \citet{brosius} analysis, a cubic-spline with a small number of nodes is used fit the line intensities to provide a smooth reconstruction of the DEM.  \citet{kashyap_drake} find a significant temperature grid of about 20-25 values (their Fig. 2) and provide 95\% confidence limits by means of an MCMC process.  The two reconstructions are shown in Fig. 2 of \citet{kashyap_drake}.  Perhaps the most noticeable differences between the two are the high values of the \citet{brosius} DEM at 1 $\times$ 10$^5$.  \citet{brosius} state that the low temperature values of the DEM are constrained by upper limits or relatively uncertain measurements of \ion{O}{iii}, \ion{C}{iv} and \ion{Mg}{v} lines but it is not stated where these line intensities come from and they are not listed in any of the tables.  The lowest temperature of importance as found by \citet{kashyap_drake} and the EMs found in the present work are about 5-6 $\times$ 10$^5$ K.  Otherwise, the DEMs of \citet{brosius} and \citet{kashyap_drake} at high temperature are mostly in agreement, taking into account the differences in approach.  The values of \citet{brosius} are smooth, as expected, while those of \citet{kashyap_drake} show some peaks in the reconstruction but the authors suggest that not all of these peaks are statistically significant. 

The emission measure distributions derived here are particular to these data sets.  For example, the observed spectra do not contain any lines that would be formed at about 10$^5$ K in the transition region.  Consequently, the emission measure at those temperature can not be determined from this data.  Since these are disk observations, the transition region is most likely to be along the line of sight.  Similarly, nothing can be said about the existence of higher temperature plasmas above about 3 $\times$ 10$^6$ K.  The observations are limited in their ability to constrain an emission measure distribution beyond a certain temperature range.
  
\citet{craig_brown} pointed out, some time ago, that the inversion of spectral line intensities to determine a continuous DEM is very unstable.  Here, empirical models are used to represent the emission measure distribution and apply Bayesian inference techniques to determine the level of detail that can be recovered from the set of observed spectral line intensities.  What is found is that the emission measure distribution in the quiet and active Sun can be determined with a maximum of 9 quantities, 4 coupled T-EM pairs and the standard deviation of the data from the model, for the set of spectral line intensities analyzed here.  Models with higher numbers of T-EM pairs are not statistically relevant.  The set of measured line intensities do not provide sufficient constraints to derive a more detailed model.

\citet{kashyap_drake} reconstruct the DEM from the same spectra analyzed here, but, by requiring a smooth solution, they are able to determine roughly twice the number of emission measure values as found in the present analysis.  In addition, they define a temperature grid for which a large number of temperatures are included.  Consequently, the number of parameters that they derive are about 4 times that of the present analysis.  \citet{hannah_kontar} employ a regularized inversion technique to determine the DEM of the active region core from EIS and SDO X-ray observations of \citet{warren_2010_eis_ar}.  Their solution is also highly detailed.  \citet{warren_bayes} employed a sparse Bayesian to examine the DEM that can be inferred from synthetic spectral line intensities created from a model DEM.  They derive 30 "weights" from 21 synthetic spectral lines and a broad-band EUV detector observation.  All three of these techniques use a smoothed or regularized solution.  This is clearly a very strong constraint on the DEM solutions.  

The use of smoothing and regularization comes from the idea that the DEM should be relatively smooth.  However, this is more of an arbitrary constraint and not one based on physical models or concepts.  If there are no physical models, then it is questionable as to what has been learned about the structure of the corona from the DEM.  The empirical models employed here only tell us what set of temperatures and emission measures are constrained by the observed spectra.  They offer only a very vague description of the solar atmosphere.  These techniques are best used in combination with physical models that make concrete predictions of the T-EM distribution.

The implication is that the derivation of a DEM or EM distribution is of little value, in and of itself.  However, the approach used here can result in some improvements in the determination of electron densities and relative elemental abundances.  In the case of determining electron densities from density-sensitive line ratios, it is necessary to select a temperature at which to calculate the predicted ratio.  This is often taken as the temperature that maximizes the {\it G(T)} function.  \citet{landi-L-function} have introduced the use of {\it L-functions} as a method of reducing the uncertainties in the determination of the temperature for calculating the line ratios.  They separate the {\it G(T)} function into two functions f$_{ij}$(N$_e$, T) and g(T) that multiply each other (see Equ. 9 of \citet{landi-L-function}).  The function g(T) contains the temperature dependence that is common to all lines of a given ion.  The temperature to be used for the calibration of the line-ratio {\it vs} density is found from the average of the temperature integrated over the product of g(T) and the DEM.  Under the current approach, there is no DEM.  In this case, one should construct a model using all of the T-EM pairs to derive the density, although this would be labor intensive.    

\citet{craig_brown} stated that one would only be able to derive a number of parameters that are much less than the number of spectral lines.  Their conclusion was based on the analysis of \citet{herring_craig} where 2 T-EM pairs, or 4 parameters, were able to account for the X-ray emission in solar flares recorded in the 7 channels of the X-ray proportional counter on the Orbiting Solar Observatory 5.  For the quiet sun analysis here, the ratio is around 1 parameter for each of 6 spectral lines observed or 1 parameter for each of 2 ions that are observed.  For the active region analysis, the ratio is around 1 parameter for each of 7 spectral lines observed or 1 parameter for each of 3 ions that are observed.  If we take our ratio of the number of parameters to the number of ions observed, then these results generally agree with the previous conclusions of \citet{craig_brown}.  They also pointed out the further constraints could be obtained from physical models but so far none have proved helpful.

\section{Conclusions}  \label{sec:conclusions}

A simple empirical model consisting of a specified number of discrete T-EM pairs has been used to explore the constraints that a set of observed spectral line intensities place on the derived emission measure distribution.  Using an MCMC process, it is determined that only 4 T-EM pairs are required to reproduce the spectral line intensities in both a quiet Sun region and an active region, for the set of observations analyzed here.  In addition, the errors in the derived parameters are determined from the MCMC posterior distributions.  In this case, they are relatively small.

The goal of this analysis has been to make an empirical determination of the ability of a set of emission line intensities to constrain the reconstruction of the emission measure distribution.  The conclusion to be drawn here is that any given set of emission line observations provide a limited amount of information with regard to the determination of the emission measure distribution.  Consequently, only a limited number of statistically meaningful parameters can be inferred, leading to reconstructions that are limited in terms of their information content, such as the temperature extent over which the distribution can be determined and the degree of detail that can be achieved.  This also would also apply to the determination of the differential emission measure.

\begin{acknowledgments}

I thank Drs. E. Landi and P. Young for helpful comments on the manuscript.  I am grateful to Dr. Jeffrey Brosius for providing the tables from \citet{brosius} in machine readable format.  This research has made use of NASA’s Astrophysics Data System.  This work has been supported by NASA grants 80NSSC21K0110 and 80NSSC21K1785.

\end{acknowledgments}

\bibliography{emd}{}
\bibliographystyle{aasjournal}

\end{document}